\documentclass[12pt]{iopart}
\eqnobysec
\usepackage{iopams} 
\usepackage{epsfig}
\usepackage{amssymb}
\begin{document}
%\documentclass[twocolumn,showpacs,eqsecnum,pra]{revtex4}
% \documentclass[preprint,showpacs,eqsecnum,pra]{revtex4}

%\topmargin 0.5cm
%\hoffset -0.85in

%\evensidemargin -0mm
%\oddsidemargin -0mm
%\usepackage{amsmath}

%\usepackage{amssymb}
%\usepackage{epsfig}
%\begin{document}
%\pagenumbering{arabic}

%\draft

\title[Hellinger distance as a measure of Gaussian discord]
{Hellinger distance as a measure of Gaussian discord}
\author{Paulina Marian$^{1,2}$
and Tudor A. Marian{$^1$}}
\address{$^1$Centre for Advanced  Quantum Physics,
Department of Physics, University of Bucharest, 
R-077125 Bucharest-M\u{a}gurele, Romania}
\address{$^2$ Department of Physical Chemistry,
University of Bucharest, Boulevard Regina Elisabeta 4-12, R-030018  Bucharest, Romania}
\ead{paulina.marian@g.unibuc.ro}
\ead{tudor.marian@g.unibuc.ro}

\date{\today}

\begin{abstract}  Especially investigated in recent years, the Gaussian discord can tentatively be quantified 
by a distance between a given two-mode Gaussian state and the set of all the zero-discord two-mode Gaussian states. However, as this set consists only of product states, such a distance captures all 
the correlations  (quantum and classical) between modes. Therefore it is merely un upper bound 
for the geometric discord, no matter which is  the employed distance. In this work we choose for this purpose the Hellinger metric that is known to have many beneficial properties recommending it as a good measure 
of quantum behaviour. In general, this metric is determined by affinity, a relative of the Uhlmann fidelity 
with which it shares many important features. As a first step of our work, the affinity of a pair of $n$-mode Gaussian states is written. Then, in the two-mode case, we succeeded in determining exactly the closest Gaussian product state and computed the Gaussian discord accordingly. The obtained general formula is remarkably simple and becomes still friendlier in the significant case of symmetric two-mode Gaussian states. We then analyze in detail two special classes of two-mode Gaussian states of theoretical and experimental interest as well: the squeezed thermal states and the mode-mixed thermal ones. The former are separable 
under a well-known threshold of squeezing, while the latter are always separable.
It is worth stressing 
that for symmetric states belonging to either of these classes, we find consistency 
between their geometric Hellinger discord and the originally defined discord in the Gaussian approach. 
At the same time, the Gaussian Hellinger discord of such a state turns out to be a reliable measure of the total amount of its cross correlations.

\end{abstract}

%\twocolumn

\maketitle
\section{Introduction}

The difference between two classically equivalent definitions of the mutual information as a measure of total correlations in a bipartite state of a composite quantum system 
stays at the origin of the quantum discord idea \cite{OZ,HV}. Such a difference arises when actions and measurements are performed on one or more subsystems. 
In the case of mixed states, quantum discord is a measure of quantumness whose relation to entanglement is not a simple one.  Evaluation of up-to-date defined degrees of correlations 
is a difficult problem both analytically and numerically. To be more specific, the entanglement of formation is defined as an optimization over all pure-state decompositions  of the given state \cite{Bennett}, while the discord emerges from an optimization over the set of all local measurements.  The computational complexity of such measures of quantum correlations in the discrete-variable systems was recently analyzed in Ref. \cite{Huang1} 
and found to increase exponentially with the dimension of the Hilbert space. That is why, despite all efforts, 
only few analytic results could be found so far even in the simplest cases. For instance, the entropic discord originally defined in Refs. \cite{OZ,HV} was investigated for two-qubit X states  in Ref. \cite{alber} but, according to Ref. \cite{Huang2}, an analytic formula of quantum discord for general two-qubit X states 
is still to be derived. 

In the continuous-variable settings, evaluation of the original discord for two-mode Gaussian states (GSs) could be accomplished 
by restricting the set of local measurements to the Gaussian ones to obtain what is now called the Gaussian discord \cite{PG,AD}.
Interestingly, for a large class of two-mode GSs it was recently proven that the Gaussian discord is an exact result, namely, the optimal local measurement defining the discord is a Gaussian 
Positive Operator-Valued Measure (POVM) \cite{PSBCL}. A review of the increasing interest, progress, and application of classical and quantum correlations quantified by quantum discord and other measures can be found in Ref. \cite{modi1}. Among these quantifiers, an important role was recently attributed to the so-called geometric measures of quantum discord \cite{modi2,DVB,Luo2}. Let us recall that quantum properties 
of the states involved in various protocols were successfully quantified using distance-type measures. 
The distance from a given state having a specific property to a reference set of states not having it can be interpreted as a quantifier of that property \cite{Hil,PVK}. When using true distances such as 
the trace  \cite{Hil},  Hilbert-Schmidt  \cite{Dod}, Bures  \cite{PTH02}, or Hellinger \cite{Luo1} ones , 
we speak about a geometric measure of that property. 
For instance, in Refs. \cite{Hil,Dod,PTH02},  geometric measures 
of non-classicality of one-mode states were defined and investigated. Alternatively, one can use other "distances" which are not true metrics, but are known to possess good distinguishability properties, 
such as the relative entropy for entanglement \cite{PVK}, discord \cite{modi2}, and non-Gaussianity \cite{PT2013} and the quantum Chernoff bound for non-classicality \cite{BGPT} and polarization \cite{GBPT}.
To efficiently apply a distance-type measure to the case of discord, a primary condition was to identify 
the set of zero-discord states 
 \cite{PHH} according to the original definition \cite{OZ,HV}.
A geometric measure was first considered by using 
the Hilbert-Schmidt metric \cite{DVB},
\begin{equation}
D_{HS} ({\hat \rho}):= \min_{\{\hat \chi\}}{[d_{HS}}({\hat \rho},\,\hat\chi)]^2,
\label{d0} 
\end{equation}
where $\{\hat \chi\}$ is the set of zero-discord states.
More recently \cite{Luo2}, a geometric discord was defined  in a slightly different manner: 
\begin{equation}
\bar D_{HS} ({\hat \rho}):=\min_{\{\hat \Pi^{(a)}\}}{[d_{HS}}
({\hat \rho},\,{\hat \Pi^{(a)}(\hat\rho}))]^2,
\label{d1} 
\end{equation}
where the minimum is taken over all the von Neumann measurements applied 
to subsystem $(a)$. The measurement operators  $\hat \Pi^{(a)}$ 
are orthogonal projectors of rank one. Being in terms of local measurements, definition\ (\ref{d1}) is therefore an inspiration of the original proposal \cite{OZ,HV}. 
The Hilbert-Schmidt geometric discord\ (\ref{d0}) was found to have a closed expression for any two-qubit state \cite{DVB}. Moreover,  Luo and Fu proved that the results of the two definitions\ (\ref{d0}) and\ (\ref{d1}) nicely coincide for any bipartite state belonging to a finite-dimensional  Hilbert space \cite{Luo2}.
As quite severe drawbacks of the Hilbert-Schmidt geometric discord  were signalled in Ref. \cite{P}, 
some other distances were recently explored to soundly define geometric measures of discord.
In Ref. \cite{Luo3} explicit formulae of the  geometric  discord\ (\ref{d1}) were given for pure states, 
as well as for $(2 \times n)$-dimensional states using the Hellinger distance. The same metric was shown 
to arise in a more general treatment of discord-type non-classical correlations based on local quantum uncertainty \cite{adesso3}. In parallel, the Bures metric was employed 
as a measure of quantum correlations  for states lying in finite-dimensional  Hilbert spaces too  \cite{SO}. When applied to two-mode Gaussian states in Refs. \cite{adesso1,adesso2}, 
the  definition\  (\ref{d1}) was modified by allowing  general  local  measurements  (POVMs). 

In the present work, we take inspiration from a recent proposal 
in Ref. \cite{Luo3} and use the Hellinger metric \cite{Luo1} 
as a geometric measure of discord for two-mode Gaussian states. 
The paper is organized as follows. In Sec. 2, we recapitulate 
some of the beneficial features of the Hellinger distance as presented 
in Refs. \cite{Luo1,Luo3}. Section 3 recalls the general structure 
of a GS. Then we write the Hellinger distance between two $n$-mode GSs 
by using some of our recent findings in Ref. \cite{PT2012}. 
In Sec. 4, we point out that any Gaussian geometric discord suffers 
from the drawback of not distinguishing between quantum and classical correlations and therefore being a Gaussian measure of the total
amount of correlations in a two-mode GS. By using the Hellinger metric,
we find analytically in Sec. 5 the nearest product GS with respect 
to a given two-mode GS. This enables us to write a closed-form 
expression of the Hellinger discord valid for an arbitrary two-mode GS. 
In Sec. 6, we show that the obtained formula turns out to be 
considerably simpler for symmetric states. Section 7 is devoted 
to the Hellinger discord for two special classes of two-mode GSs which are interesting for theory and experiment as well: the squeezed thermal states and the mode-mixed thermal ones. In Sec. 8, the agreement between the Hellinger measure of all the correlations and the quantum  mutual information is illustrated within the Gaussian approach.  We conclude with a summary of our results. Finally,
in the Appendix, for some types of two-mode GSs, we express the covariance matrix of the square-root state in terms of that 
of the state itself.

\section{Fidelity and affinity}

The statistical overlap between two probability  distributions 
$f(s)$ and $g(s)$,
\begin{equation}
{\cal A}(f,g):=\sum_{s}\sqrt{f(s) g(s)},
\label{so}
\end{equation}
is widely used in statistical physics and interpreted as a measure of classical distinguishability \cite{B}. Indeed, 
Eq.\ (\ref{so}) is one of the R\'enyi overlaps  which are distinguishability measures in their own right \cite{Fuchs}. In the quantum scenario, one usually considers a general measurement (POVM), i. e., a set 
of non-negative operators ${\{\hat E_b\}}$ which is complete on the Hilbert space ${\cal H}$, or, 
in other words, is a resolution of the identity:  
${\sum_{b} \hat E_b}=\hat I$. 
The subscript $b$ indexes the possible outcomes of the measurement whose probabilities in the quantum states $\hat \rho$ 
and $\hat \sigma$ are, respectively, 
$p_{\hat \rho}(b)={\rm Tr}(\hat \rho \hat E_b)$ 
and $p_{\hat \sigma}(b)={\rm Tr}(\hat \sigma \hat E_b).$  

An indicator of the closeness of the two quantum states 
via measurements can be built by associating to any POVM  
the statistical overlap\ (\ref{so}) of the probability 
distributions $p_{\hat \rho}(b)$ and $p_{\hat \sigma}(b)$:
\begin{equation}
{\cal A}_{\{\hat E_b\}}(p_{\hat \rho}, p_{\hat \sigma})
:=\sum_b \sqrt{p_{\hat \rho}(b)p_{\hat \sigma}(b)}.
\label{ca}
\end{equation}
An important theorem proven in Refs. \cite{Fuchs,FC,barn1,barn2} states that the minimal overlap\ (\ref{ca}) over all quantum measurements is equal to the square root of fidelity:
\begin{equation}
\min_{\{\hat E_b\}}[{\cal A}_{\{\hat E_b\}}(p_{\hat \rho},
p_{\hat \sigma})]=\sqrt{{\cal F}(\hat \rho, \hat \sigma)}.
\label{fi1}
\end{equation}
Nowadays a widely accepted figure of merit in quantum information science, fidelity has been introduced by Uhlmann 
in a quantum-mechanical framework \cite{Uhl}. It is related to 
the Bures distance \cite{Bures},
\begin{equation}
[d_{B}(\hat \rho,\hat\sigma)]^2
:=2-2\sqrt{{\cal F}(\hat\rho,\hat\sigma)}, 
\label{db}
\end{equation}
and to the statistical distance \cite{BC}. Uhlmann derived 
the explicit expression of fidelity in terms of density operators \cite{Uhl} : 
\begin{equation}
\sqrt{{\cal F}(\hat \rho, \hat \sigma)}={\rm Tr}\left[ \left(
\sqrt{\hat \sigma}\,\hat \rho\,\sqrt{\hat \sigma}\right)^{\frac{1}{2}}\right]
=||\sqrt{\hat \rho}\,\sqrt{\hat \sigma}||_1.
\label{fi2}
\end{equation}
Here $||\hat B||_1:={\rm Tr}|\hat B|$ denotes the trace norm. 
A friendly review of its main properties was written
by Jozsa \cite{Jozsa}.

Why is theorem\ (\ref{fi1}) so interesting? First, it tells us 
that fidelity cannot decrease under any POVM. This feature is important in defining fidelity-based distance-type measures of various quantum properties. The most prominent example is 
the Bures metric\ (\ref{db}) which is therefore contractive.
Second, Eq.\ (\ref{fi1}) exhibits the square root of fidelity 
as a classically defined statistical overlap\ (\ref{so})
in the quantum-measurement version. 

However, an appropriate quantum analogue 
of the overlap\ (\ref{so}) of two probability distributions has recently been pointed out in Refs. \cite{Luo1,Luo3}. This is the affinity of two quantum states $\hat\rho$ and $\hat\sigma$:
\begin{equation}
{\cal A}(\hat\rho,\,\hat\sigma):={\rm Tr}(\sqrt{\hat\rho}\, 
\sqrt{\hat\sigma}). 
\label{aff} 
\end{equation}
Affinity is a non-negative functional which is symmetric with
respect to $\hat\rho$ and $\hat\sigma$ and is connected
to the Hellinger distance between these states \cite{Luo1}:
\begin{equation}
[d_H(\hat \rho,\hat \sigma)]^2:={\rm Tr}\left[ \left( 
\sqrt{{\hat \rho}}-\sqrt{{\hat \sigma}}\right)^2 \right]
=2-2 {\cal A}(\hat \rho,\hat \sigma).
\label{hd} 
\end{equation}
Note that we employ as Hellinger distance the square root $d_H$ 
of the original one introduced in Ref. \cite{Luo1} on probabilistic grounds, because $d_H$ fulfills the triangle inequality \cite{Mend} and therefore is a true metric.

Affinity can be considered a closeness measure in its own right. 
Its aspect indicates a close relation with the square root of the Uhlmann fidelity\ (\ref{fi2}). 
Indeed, the square root of fidelity as well as affinity possess 
some convenient features that ensure both the Bures and Hellinger 
metrics, Eqs.\ (\ref{db}) and\ (\ref{hd}), respectively, to be good candidates for quantifying properties such as quantum correlations. 
We list them as follows \cite{Fuchs,Luo1}.
\begin{enumerate}
\item ${\cal A}(\hat\rho,\,\hat\sigma)\leqq 1$.\\ 
Affinity reaches unity if and only if the quantum states 
$\hat\rho$ and $\hat\sigma$ coincide.
\item {Invariance under unitary transformations:}
$${\cal A}(\hat U \hat \rho \hat U^{\dag}, \hat U \hat \sigma
\hat U^{\dag})={\cal A}(\hat \rho, \hat \sigma).$$
\item {Multiplicativity:} 
$${\cal A} (\hat\rho_1\otimes\hat\rho_2, \hat\sigma_1\otimes\hat\sigma_2)={\cal A}(\hat \rho_1,\hat \sigma_1)
{\cal A}(\hat \rho_2,\hat \sigma_2).$$
\item {Joint concavity with respect to both arguments:} 
$${\cal A}\left( \sum_j\lambda_j\hat \rho_j, 
\sum_j\lambda_j\hat \sigma_j \right) \geqq \sum_j\lambda_j
{\cal A}(\hat \rho_j,\hat \sigma_j), \qquad \left( \lambda_j>0, \quad \sum_j\lambda_j=1 \right).$$
This property obviously implies the concavity of affinity in any
of its two arguments.
\item The inequality 
$${\cal A}(\hat\rho, \hat\sigma)\leqq 
\sqrt{{\cal F}(\hat \rho, \hat \sigma)}$$
holds \cite{Luo1} and saturates if and only if the states commute:
$${\cal A}(\hat\rho, \hat\sigma)
=\sqrt{{\cal F}(\hat \rho, \hat \sigma)}\;\Longleftrightarrow \;
[\hat\rho,\,\hat\sigma]=\hat 0.$$
\end{enumerate}
We emphasize that affinity can be employed in an equivalent way 
with another figure of merit in discriminating between two quantum states, namely, their trace distance. This property was found 
long ago by Holevo, who proved the following pair of inequalities 
\cite{Hol}:
\begin{equation}
1-{\cal A}(\hat\rho, \hat\sigma)\leqq T(\hat\rho, \hat\sigma) 
\leqq \sqrt{1-[{\cal A}(\hat\rho, \hat\sigma)]^2}.
\label{4}
\end{equation}
Here $T(\hat\rho,\hat\sigma):=\frac{1}{2}||\hat\rho
-\hat\sigma||_1$ is the trace distance between the states 
$\hat\rho$ and $\hat\sigma $. More recently, Fuchs and 
van de Graaf derived similar inequalities for the square root 
of fidelity \cite{FG}:
\begin{equation} 
1-\sqrt{{\cal F}(\hat\rho, \hat\sigma)}\leqq T(\hat\rho,
\hat\sigma) \leqq \sqrt{1-{\cal F}(\hat\rho, \hat\sigma)}.
\label{4a}
\end{equation}
The trace metric is a particularly important distinguishability 
measure owing to its connection with the probability of error 
between two quantum states \cite{FG}. Recall that it was used 
by Hillery to formulate the first proposal of a distance-type 
measure of non-classicality in Ref. \cite{Hil}. Despite of still being considered as a sensitive distance-type measure of any quantum property \cite{Breuer}, the trace metric remains difficult to deal with analytically. 
In turn,  Eqs.\ (\ref{4}) and\ (\ref{4a}) show that both the Bures metric,
$1-\sqrt{{\cal F}(\hat\rho,\hat\sigma)}$, 
and the Hellinger metric, $1-{\cal A}(\hat\rho,\hat\sigma)$, compare closely to the trace metric as measures 
of distinguishability. In contrast to fidelity, affinity presents the advantage of being easily computable 
for many classes of states.

\section{Affinity of two $n$-mode Gaussian states}

The Gaussian states of the quantum radiation field are important resources in many quantum information protocols \cite{BL,WPGCRSL,QC3}.
In particular, the two-mode ones are experimentally quite accessible and constitute a perfect test bed for studying 
all kind of correlations between modes. 
It appears to the present authors that the affinity of two arbitrary 
$n$-mode GSs was first derived  by Holevo in Ref.\cite{Hol} 
within the formalism of the $C^*$-algebra of commutation relations. 
Here we recover it by using our recent findings on the products of two Gaussian operators in Ref. \cite{PT2012}, where we have also synthesized
some of the notions and notations employed for GSs. From now on, 
we drop the subscript $G$ expressing the Gaussian character of the operators we deal with. 

Let us consider an arbitrary $n$-mode GS ${\hat \rho}$. We find it convenient to introduce a $2n$-dimensional row vector $(\hat u)^T$  consisting of the canonical quadrature operators of all the modes:
\begin{equation}
(\hat u)^T:=(\hat q_1,\; \hat p_1,\;\dots,\; \hat q_n,\;\hat p_n),
\label{hatu}
\end{equation}
where $\hat{q}_j,\; \hat{p}_j$ are the quadrature operators 
of the $j^{\mathrm{th}}$ field mode. Further, we employ a related 
vector $u \in \mathbb{R}^{2n}$ whose components are arbitrary 
eigenvalues of the quadrature operators of the modes:
\begin{equation}
u^T:=(q_1,\; p_1,\;\dots,\; q_n,\; p_n).
\label{Tu}
\end{equation}
In particular, let us denote $v \in \mathbb{R}^{2n}$ the vector
consisting of the expectation values of the quadrature operators
in the chosen state ${\hat \rho}$:
$v:={\langle{\hat u}\rangle}_{\hat{\rho}}.$
The second-order moments of the deviations from the means 
of the canonical quadrature operators are collected as entries 
of the symmetric covariance matrix (CM) 
${\mathcal V}\in M_{2n}(\mathbb{R}),$ which is positive definite: ${\mathcal V}>0$. Recall that the characteristic function 
$\chi (u)$ of the GS $\hat{\rho}$ is determined only by its first- and second-order moments of the canonical quadrature operators and has a specific exponential form:
\begin{equation}
\chi (u):={\rm Tr}\left[ \hat{D}(u) \hat{\rho}\right]
=\exp\left(-\frac{1}{2}\,u^{T}{\mathcal V}\,u-iv^{T}u\right).
\label{CF}
\end{equation}
In Eq.\ (\ref{CF}), $\hat{D}(u)$ stands for an  $n$-mode Weyl displacement operator, whose factors are single-mode Weyl shifting operators:
\begin{equation}
\hat{D}(u):=\bigotimes_{j=1}^{n}\hat{D}_j(q_j,\, p_j), \qquad
\hat{D}_j(q_j,\, p_j):
=\exp \left[-i\left(q_j\hat{p}_j-p_j\hat{q}_j\right)\right].
\label{D}
\end{equation}

According to Williamson's theorem \cite{W}, the CM ${\mathcal V}$ is congruent via a symplectic matrix $S \in Sp(2n, \mathbb{R})$ to a diagonal matrix:
\begin{equation}
{\cal V}=S\left[\bigoplus_{j=1}^{n}\,\left({\kappa}_j\,\sigma_0
\right) \right]S^T, \qquad \det({\cal V})
=\prod_{j=1}^{n}({\kappa}_j)^2.  
\label{W}
\end{equation}
In Eq.\ (\ref{W}), the positive numbers ${\kappa}_j, \;\; 
(j=1,2,\dots, n),$ are the symplectic eigenvalues of the CM 
${\cal V}$, and $\sigma_0$ denotes the $2\times 2$ identity 
matrix. 
We take advantage of the standard matrix $J$ of the symplectic form on $\mathbb{R}^{2n},$ 
\begin{eqnarray}
J:=\bigoplus_{k=1}^{n}\,J_k, \quad J_k:=\left( 
\begin{array}{cc}
0  & 1\\ -1 & 0
\end{array} 
\right), \quad (k=1,2,\dots, n),
\label{J}
\end{eqnarray}
to write the Robertson-Schr\"odinger uncertainty relation 
in two equivalent ways:
\begin{equation}
{\cal V}+\frac{i}{2}J  \geqq 0\;\;\Longleftrightarrow \;\; {\kappa}_j \geqq \frac{1}{2}, \qquad (j=1,2,\dots, n).
\label{R&S}
\end{equation}

It is worth emphasizing the Hilbert-space counterpart of the Williamson transformation\ (\ref{W}), namely, 
the statement that any $n$-mode GS ${\hat \rho}$ is unitarily similar to an $n$-mode thermal state (TS) 
${\hat \rho}_T$:
\begin{equation}
{\hat \rho}=\hat D(v)\hat U(S)\hat \rho_T\hat U^{\dag}(S)\hat D^{\dag}(v).
\label{u}
\end{equation}
The mapping  $\hat U(S)$ is called the metaplectic representation of the symplectic group $Sp(2n, \mathbb{R})$. 
Obviously, the density operators ${\hat \rho}$ and $\hat \rho_T$ 
have the same discrete spectrum consisting of non-degenerate 
and strictly positive eigenvalues. These are determined
by the symplectic eigenvalues of the CM ${\cal V}$, 
\begin{equation}
\kappa_j =\bar{n}_j+\frac{1}{2}, \qquad (j=1,2,\dots, n),
\label{kappaj}
\end{equation}
where $\bar{n}_j$ are the mean photon occupancies of the modes in 
the TS $\hat \rho_T$. 

For further convenience, we define a square-root state as
\begin{equation}
\hat \rho_{sr}:=\left( {\rm Tr} \sqrt{\hat \rho}\right)^{-1}
\sqrt{\hat \rho}.
\label{sr}
\end{equation}
Let us consider the square-root state $(\hat \rho_T)_{sr}$, 
which is itself an $n$-mode TS. Its CM has the symplectic eigenvalues specified in Ref. \cite{PT2012}: 
\begin{equation}
\tilde{\kappa_j}=\kappa_j+\sqrt{{\left({\kappa}_j \right)}^2
-\frac{1}{4}}, \qquad (j=1,2,\dots, n).
\label{srsymp}
\end{equation} 
Note that Eq.\ (\ref{u}) implies the transformation formula
\begin{equation}
\hat \rho_{sr}=\hat D(v)\hat U(S)(\hat \rho_T)_{sr}\,\hat U^{\dag}(S)
\hat D^{\dag}(v).
\label{usr}
\end{equation}
This means that $\hat \rho_{sr}$ is an $n$-mode GS whose CM is obtained from the CM of the state ${\hat \rho}$ just by replacing 
the symplectic eigenvalues $\kappa_j$ with the corresponding ones 
$\tilde{\kappa_j}$, Eq.\ (\ref{srsymp}). 

The product of two Gaussian density operators, 
$\hat{\mathcal B}:=\hat{\rho}^{\prime}\hat{\rho}^{\prime \prime},$
is largely investigated in Ref. \cite{PT2012} using as a main tool 
the characteristic function\ (\ref{CF}) of the GS $\hat{\rho}$.
It turns out that $\hat{\mathcal B}$ is an $n$-mode Gaussian operator whose weight function in the Weyl expansion is of the type\ (\ref{CF}) 
up to a prefactor \cite{PT2012}. Its symmetric matrix 
${\mathcal F}\in M_{2n}(\mathbb{C})$ is positive definite and is determined by the following composition rule \cite{PT2012}:
\begin{equation}
{\mathcal F}=-\frac{i}{2}J+\left({\mathcal V}^{\prime \prime}
+\frac{i}{2}J \right)
\left({\mathcal V}^{\prime}+{\mathcal V}^{\prime \prime}\right)^{-1}
\left({\mathcal V}^{\prime}+\frac{i}{2}J\right),
\label{rhorho}
\end{equation}
Further, in Ref. \cite{PT2012}, we have written the trace of the operator $\hat{\mathcal B}$ in terms of the CMs ${\mathcal V}^{\prime}$ and ${\mathcal V}^{\prime\prime}$ of the two factors, and the relative average $n$-mode displacement 
$\delta v:=v^{\prime}-v^{\prime \prime}
=\langle{\hat u}\rangle_{{\hat{\rho}}^{\prime}}
-\langle{\hat u}\rangle_{{\hat{\rho}}^{\prime \prime}}$: 
 \begin{eqnarray}
{\rm Tr}\,(\hat{{\mathcal B}})=\left[\det\left({\cal V}^{\prime}
+{\cal V}^{\prime\prime}\right)\right]^{-\frac{1}{2}} 
\, \exp{\left[-\frac{1}{2}\left(\delta v \right)^T \left({\cal V}^{\prime}+{\cal V}^{\prime\prime}\right)^{-1} 
\delta v \right]}>0. 
\label{TR_HS} 
\end{eqnarray}

According to Eq.\ (\ref{aff}), our evaluation of the affinity of 
two multi-mode GSs involves square roots of Gaussian density operators. 
 Let us denote by 
 $\tilde{\mathcal V}^{\prime}$ and $\tilde{\mathcal V}^{\prime\prime}$ the CMs of the GSs $(\hat{\rho}^{\prime})_{sr}$ and 
$(\hat{\rho}^{\prime\prime})_{sr}$, respectively.
Taking account of Eqs.\ (\ref{sr}),\ (\ref{usr}), and\ (\ref{TR_HS}), 
we get the affinity 
\begin{eqnarray}
&{\cal A}(\hat\rho^{\prime}, \hat\rho^{\prime\prime}):={\rm Tr}\left( \sqrt{\hat{\rho}^{\prime}}\,
\sqrt{\hat{\rho}^{\prime\prime}}\right)=
\frac{{\rm Tr}\left(\sqrt{\hat{\rho}^{\prime}}\right){\rm Tr}\left(\sqrt{\hat{\rho}^{\prime\prime}}\right)}
{\left[\det\left(\tilde{\mathcal V}^{\prime}
+\tilde{\mathcal V}^{\prime\prime}\right)\right]^{\frac{1}{2}}} 
\nonumber\\
&\times \exp{\left[-\frac{1}{2}\left(\delta v \right)^T 
\left(\tilde{\mathcal V}^{\prime}
+\tilde{\mathcal V}^{\prime\prime}\right)^{-1}\delta v \right]}. 
\label{sqTR} 
\end{eqnarray}
In the particular case $\hat{\rho}^{\prime}
=\hat{\rho}^{\prime\prime}=:\hat{\rho},$  Eq.\ (\ref{sqTR}) reduces 
to the identity
\begin{equation}
{\rm Tr}\left(\sqrt{\hat{\rho}}\right)={\left[\det\left(2 \tilde{\mathcal V}\right)\right]^{\frac{1}{4}}}.
\label{sqTR1} 
\end{equation}
Besides, Eq.\ (\ref{rhorho}) becomes a relationship between the CMs 
${\mathcal V}$ and $\tilde{\mathcal V}$ of the $n$-mode GSs 
$\hat{\rho}$ and $\hat{\rho}_{sr}$, respectively:
\begin{equation}
{\cal V}=\frac{1}{2}\left(\tilde{\cal V}
-\frac{1}{4}J{\tilde{\cal V}}^{-1}J \right).
\label{tildeV}
\end{equation}
 We now are ready to write down the affinity of two multi-mode GSs:
\begin{eqnarray}
{\cal A}(\hat\rho^{\prime}, \hat\rho^{\prime\prime})=\frac{2^{n}
\left[ \det(\tilde{\mathcal V}^{\prime})\,\det(\tilde{\mathcal V}^{\prime\prime})\right]^{\frac{1}{4}}}{\left[\det\left(\tilde{\mathcal V}^{\prime}
+\tilde{\mathcal V}^{\prime\prime}\right)\right]^{\frac{1}{2}}} 
\, \exp{\left[ -\frac{1}{2}\left(\delta v \right)^T 
\left( \tilde{\mathcal V}^{\prime}+\tilde{\mathcal V}^{\prime\prime}\right)^{-1}\delta v \right]}. 
\label{sqTR2} 
\end{eqnarray}
Remark that the affinity\ (\ref{sqTR2}) is written in terms of the CMs 
$\tilde{\mathcal V}^{\prime}$ and $\tilde{\mathcal V}^{\prime\prime}$ 
of the square-root states $(\hat{\rho}^{\prime})_{sr}$ and 
$(\hat{\rho}^{\prime\prime})_{sr}$, respectively.

\section{Significance of any Gaussian geometric discord}

The reason to define distance-type quantifiers of various 
properties is in general given by the presumably simpler way of their evaluation. We are expecting that the results obtained 
by a distance-type measure to be consistent with the original meaning of that property, usually based on complicated 
extremization procedures. If not shown by explicit analytic 
results, consistency could be checked by observing the behaviour 
of the competing measures under some accepted  requirements formulated adequately for any quantum property.

Interestingly, when analyzing the consistency between different treatments of Gaussian discord, we are from the very beginning 
in a privileged position to draw some conclusions. This happens because the originally defined discord \cite{OZ,HV} 
for two-mode GSs has been calculated under the approach 
of limiting  the set of all invoked one-party measurements 
to the Gaussian ones \cite{PG,AD}. We were thus provided with 
an analytic formula of the Gaussian discord in terms of one-mode von Neumann entropies. Moreover, according to Ref. \cite{PSBCL}, at least for the special two-mode GSs analyzed in the present paper, the Gaussian discord is {\em the} discord. 
A counterintuitive result has also emerged:  
it has been found that the only zero-discord GSs are the product ones \cite{PG,AD}. Even the separable GSs identified with Simon's criterion \cite{Si} do contain a good amount of quantum correlations measured by their discord. This finding has serious consequences on the interpretation of a geometric Gaussian discord defined
in Eq.\ (\ref{d0}), regardless of the distance we use.
Indeed, following Ref.\cite{DVB}, the definition of a geometric Gaussian discord for the two-mode GS 
${\hat \rho}$ could be, similarly to Eq.\ (\ref{d0}) :
\begin{equation} 
D_G ({\hat \rho})\sim \min_{{\hat \sigma} \in {\cal G}_P}
[{d}({\hat \rho},\,{\hat \sigma})]^2.
\label{gd} 
\end{equation}
In Eq.\ (\ref{gd}),  ${\cal G}_P$ is the set of all two-mode 
product GSs ${\hat \sigma}$ and $d$ is any distance having the required properties of discriminating among quantum states \cite{PVK}. 
As the reference set of zero-discord states coincides with that of uncorrelated (product) states, in fact, Eq.\ (\ref{gd}) gives a measure of {\em all} the Gaussian correlations. Therefore, while according to the recent results of Ref.\cite{PSBCL} the originally defined discord could be Gaussian, all geometric Gaussian measures of discord appear to be just upper bounds for the corresponding geometric discord. They quantify all the correlations in a two-mode GS, both quantum and classical, being thus geometric analogues of the quantum mutual information, 
\begin{equation}
 {\cal I}(\hat {\rho}_{AB}):={\cal S}(\hat \rho_A)+{\cal S}(\hat \rho_B)
 -{\cal S}(\hat \rho_{AB}).
\label{MI}
\end{equation}
In Eq.\ (\ref{MI}), ${\cal S}(\hat \rho)$ denotes the von Neumann entropy of the state $\hat \rho$: ${\cal S}(\hat \rho):
=-{\rm Tr}[\hat \rho \ln (\hat \rho)].$ 
In general, ${\cal I}(\hat {\rho}_{AB})$ is known to measure the total amount of correlations between the subsystems $A$ and $B$ in any given bipartite state $\hat \rho_{AB}$. 
It appears that  we cannot separate the quantum correlations from their classical counterparts when using a distance-type Gaussian measure for the quantum discord. We have thus to admit that the closest zero-discord state to a Gaussian one is not Gaussian and this happens for {\em any} distance used as a quantifier of quantum discord.

As far as we know, a geometric discord for two-mode Gaussian states  
was evaluated using the  Hilbert-Schmidt metric in Ref. \cite{adesso1}, and its {\em rescaled} version in the recent Ref. \cite{adesso2}. 
Nevertheless, we find it instructive to take advantage of the beneficial properties of affinity and define a Gaussian geometric discord 
of the type\ (\ref{d0}) in terms of the Hellinger distance between 
an arbitrary two-mode GS $\hat \rho $ and the whole set ${\cal G}_P$ 
of the zero-discord GSs, i. e., the product ones:
\begin{equation} 
D_H ({\hat \rho}):=\min_{{\hat \sigma} 
\in {\cal G}_P}\frac{1}{2}[d_H ({\hat \rho},{\hat \sigma})]^2
= 1-\max_{{\hat \sigma} \in {\cal G}_P}{\cal A}({\hat \rho},
{\hat \sigma}).
\label{gdh} 
\end{equation}

To perform the optimization\ (\ref{gdh}), we shall use 
the relationship between the CMs  
$\tilde{\mathcal V}_p$ and ${\mathcal V}_p$ in the case of a product state. This is considered in the Appendix together with a more general example.  We there write the entries of the CM $\tilde{\mathcal V}$ corresponding to a scaled standard-form CM ${\mathcal V}$ and find it 
to be in a scaled standard form as well.  

\section{Gaussian discord with the Hellinger distance}

\subsection{The closest Gaussian product state}

According to definition\ (\ref{gdh}), we now address 
the following question: which  product state belonging
to the set ${\cal G}_P$ has the maximal affinity with respect to
a given two-mode GS $\hat \rho$ ?  Let us choose a state $\hat \rho$ whose CM ${\mathcal V}$ has a scaled standard form,
\begin{eqnarray}
{\mathcal V}=\left(
\begin{array} {cc}
\mathcal V_{1}       & {\mathcal C}\\ 
{\mathcal C} & \mathcal V_{2} 
\end{array}
\right),
\label{st1}
\end{eqnarray}
i. e., it is partitioned into the following $2\times 2$ scaled 
diagonal submatrices: 
\begin{eqnarray} 
&\mathcal V_{j}=\left(
\begin{array}  {cc}
b_j s_j & 0\\ 
0 & b_js^{-1}_j
\end{array}
\right),
\quad {\mathcal C}=\left(
\begin{array}  {cc}
c\sqrt{s_1s_2} & 0\\ 
0 & d (s_1s_2)^{-\frac{1}{2}}
\end{array}
\right), \nonumber\\
&\left(j=1,2,\;\; b_{j}\geqq \frac{1}{2},
\;\; c\geqq |d|\right). 
\label{st2}
\end{eqnarray}
In Eq.\ (\ref{st2}),  $s_j$ are one-mode squeeze factors.
It has been proven in the Appendix that the CM 
$\tilde{\mathcal V}$ of the square-root state\ (\ref{sr}) has
a similar scaled standard form\ (\ref{st1})--\ (\ref{st2}), 
with entries specified by Eqs.\ (\ref{tildeb}) 
and\ (\ref{tildeb1}).
Let us examine the affinity\ (\ref{sqTR2}) between the given GS 
$\hat \rho$ whose CM is $\mathcal V$, 
Eqs.\ (\ref{st1})--\ (\ref{st2}), and a product state 
$\hat\sigma$ whose CM is 
${\mathcal V}_p$, Eqs.\ (\ref{st3})--\ (\ref{prod}). 
It is obvious that the closest product state, hereafter denoted 
by $\hat\sigma_*$, has to maximize the exponential 
in Eq.\ (\ref{sqTR2}), so that it should have the same displacement parameters as $\hat \rho$. We are thus left to perform the maximization of the undisplaced affinity
\begin{eqnarray}
&{\cal A}_1(\hat\rho, \hat\sigma)
=\frac{4\left[\det(\tilde{\mathcal V})
\det(\tilde{\mathcal V}_p)\right]^{\frac{1}{4}}}{\left[ \det\left(
\tilde{\mathcal V}+\tilde{\mathcal V}_p\right)\right]^{\frac{1}{2}}}, 
\label{a1}
\end{eqnarray}
 with respect to the one-mode squeeze parameters $\varphi_j,r_j$, 
and the symplectic eigenvalues $\eta_j$. We routinely find that  
$\det(\tilde{\mathcal V}+\tilde{\mathcal V}_p)$ is minimal 
at $\varphi_j=0,\; (j=1,2)$, so that our problem 
is reformulated as the maximization of the function
\begin{eqnarray}
&{\cal A}_2(\hat\rho, \hat\sigma):=4\sqrt{\frac{{\tilde \kappa}_1 {\tilde \kappa}_2 {\tilde \eta}_1{\tilde \eta}_2}{\delta_1\delta_2}},
\label{a2}
\end{eqnarray}
with respect to the variables $r_j,\,\tilde \eta_j$. 
In Eq.\ (\ref{a2}), use is made of the resolution into factors
$\,\det(\tilde{\mathcal V}+\tilde{\mathcal V}_p)=\delta_1 \delta_2,$ 
with:
\begin{eqnarray}
&\delta_1:=(\tilde b_1 \tilde s_1+\tilde \eta_1 {\rm e}^{2 r_1})(\tilde b_2 \tilde s_2+\tilde \eta_2 {\rm e}^{2 r_2})-{\tilde c}^2 \tilde s_1 \tilde s_2,\nonumber \\
&\delta_2:=\left( \frac{\tilde b_1}{\tilde s_1}+\tilde \eta_1 
{\rm e}^{-2 r_1}\right) \left(\frac{\tilde b_2}{ \tilde s_2}
+\tilde \eta_2 {\rm e}^{-2 r_2}\right) -\frac{{\tilde d}^2}
{\tilde s_1 \tilde s_2}.
\label{del}
\end{eqnarray}
We finally list four conditions to be  fulfilled by the variables
$r_1, r_2, \tilde \eta_1, \tilde \eta_2$ at the maximum point 
of the affinity\ (\ref{a2}): 
\begin{eqnarray}
&(\tilde b_1 \tilde s_1+\tilde \eta_1 {\rm e}^{2 r_1})(\tilde b_2 \tilde s_2-\tilde \eta_2 {\rm e}^{2 r_2})-{\tilde c}^2 \tilde s_1 \tilde s_2=0,
\nonumber \\
&(\tilde b_1 \tilde s_1-\tilde \eta_1 {\rm e}^{2 r_1})(\tilde b_2 \tilde s_2+\tilde \eta_2 {\rm e}^{2 r_2})-{\tilde c}^2 \tilde s_1 \tilde s_2=0,
\nonumber \\
&\left( \frac{\tilde b_1}{\tilde s_1}+\tilde \eta_1 {\rm e}^{-2 r_1}\right) \left( \frac{\tilde b_2}{ \tilde s_2}-\tilde \eta_2 {\rm e}^{-2 r_2}\right) -\frac{{\tilde d}^2}{\tilde s_1 \tilde s_2}=0,\nonumber \\
&\left(\frac{\tilde b_1}{\tilde s_1}-\tilde \eta_1 {\rm e}^{-2 r_1}
\right) \left( \frac{\tilde b_2}{ \tilde s_2}+\tilde \eta_2 
{\rm e}^{-2 r_2}\right) -\frac{{\tilde d}^2}{\tilde s_1 \tilde s_2}=0.
\label{max2}
\end{eqnarray}
The closest product two-mode state $\hat \sigma_* $ to the given state $\hat \rho$ is determined by the following parameters 
arising from  Eq.\ (\ref{max2}):   
\begin{eqnarray}
&(\tilde \eta_1)_*=\sqrt{\frac{\tilde b_1}{\tilde b_2}\tilde \kappa_1 \tilde \kappa_2}, \quad
(\tilde \eta_2)_*=\sqrt{\frac{\tilde b_2}{\tilde b_1}\tilde \kappa_1 \tilde \kappa_2},\nonumber\\
&{\rm e}^{2(r_j)_*}=\tilde s_j\left[\frac{\tilde b_1\tilde b_2-{\tilde c}^2}{\tilde b_1\tilde b_2-{\tilde d}^2}\right]^{\frac{1}{4}},
\quad (\varphi_j)_*=0,\quad (j=1,2).
\label{max3}
\end{eqnarray}
Although  $\hat \sigma_* $ is the nearest product state 
to the state $\hat \rho$ when using the Hellinger metric, we find it 
convenient to determine it in terms of the parameters 
of the square-root state $\hat \rho_{sr}$, Eq.\ (\ref{sr}).

Some remarks concerning our results\ (\ref{max3}) are now at hand. 
\begin{enumerate}
\item Equations\ (\ref{max3}) and\ (\ref{sr}) show that the GSs 
$\hat \sigma_*$ and $\hat \rho_{sr}$ have the same purity:
\begin{equation}
{\rm Tr}\left[ (\hat \sigma_*)^2 \right]
={\rm Tr}\left[ (\hat \rho_{sr})^2 \right]
=(4\tilde \kappa_1 \tilde \kappa_2)^{-1}.
\end{equation}
\item The closest GS  $\hat \sigma_* $ is a product of one-mode squeezed thermal states for an arbitrary  given undisplaced GS 
$\hat \rho$, except for any state with $|d|=c,$ whose closest product state is a two-mode thermal one.
\item For any undisplaced and scaled pure GS, the closest state 
$\hat \sigma_*$ is also a pure state, namely, the product of 
one-mode squeezed vacuum states  having the squeeze parameters $(r_j)_*=\ln s_j,\; (j=1,2),$ specified by Eq.\ (\ref{max3}).
Recall that, apart from the vacuum state, the only unscaled and 
unshifted pure two-mode GSs are the two-mode squeezed vacuum states. 
They are characterized by the following properties of their
standard-form parameters \cite{PTH2003,PT2008}: 
\begin{equation}
b_1=b_2=:b>\frac{1}{2},\quad c=-d>0,\quad 
b^2-c^2=\frac{1}{4}.
\label{sfpSVS}
\end{equation}
\end{enumerate}

\subsection{Explicit result}

The maximal affinity can straightforwardly be written in terms 
of the parameters of the square-root state $\hat \rho_{sr}$, 
Eq.\ (\ref{sr}), via Eqs.\ (\ref{max3}). We get first
\begin{eqnarray}
&(\delta_1)_*=2 \tilde s_1 \tilde s_2 \sqrt{\tilde b_1 \tilde b_2
-{\tilde c}^2}\left[\sqrt{\tilde b_1\tilde b_2}
+\sqrt{\tilde b_1\tilde b_2-{\tilde c}^2}\right],\nonumber\\
&(\delta_2)_*=2 \frac{1}{\tilde s_1 \tilde s_2} \sqrt{\tilde b_1\tilde b_2-{\tilde d}^2}\left[\sqrt{\tilde b_1\tilde b_2}
+\sqrt{\tilde b_1\tilde b_2-{\tilde d}^2}\right],
\end{eqnarray} 
which yields, via Eq.\ (\ref{a2}), the maximal affinity 
\begin{eqnarray}
&{\mathcal A}(\hat \rho):={\mathcal A}(\hat \rho,\, \hat \sigma_*)
\nonumber\\
&=\left[\frac{4\sqrt{\det \tilde{\mathcal V}}}{\left(\sqrt{\tilde b_1\tilde b_2}+\sqrt{\tilde b_1\tilde b_2
-{\tilde c}^2}\right) \left(\sqrt{\tilde b_1\tilde b_2}
+\sqrt{\tilde b_1\tilde b_2-{\tilde d}^2}\right)}\right]
^{\frac{1}{2}}.
\label{MA}
\end{eqnarray} 
As expected from the properties of the Hellinger distance, 
the Gaussian discord\ (\ref{gdh}) evaluated via Eq.\ (\ref{MA}) does not depend on the local squeezing factors. It has a remarkably simple analytic aspect in terms of the entries of the CM 
$\tilde {\cal V}$. This nice formula hides a rather complicated structure when expressed with the entries of the CM ${\cal V}$. 
For instance, it does not display a dependence of the sign of the parameter $\tilde d$.  However, as we shall see and comment later,  application of the transformation 
relations\ (\ref{tildeb})--\ (\ref{tildeb4})  gives us distinct
expressions for the maximal affinity of the states whose CMs 
differ only by the sign of $d$. 

A tedious calculation of $\delta_{1*}$ and $\delta_{2*}$ by inserting 
the correspondence rules\ (\ref{tildeb}) leads 
to a convenient form of the maximal affinity\ (\ref{MA}):
\begin{eqnarray}
\mathcal A (\hat \rho)=\frac{2 {\cal K}(\det {\cal V})^{\frac{1}{4}}}
{\left[{\cal K}^2 \sqrt{\det {\cal V}}
+{\cal K}(\det {\cal V})^{\frac{1}{4}} \sqrt{\cal Q}
+\sqrt{{\cal B}_1 {\cal B}_2}\right]^{\frac{1}{2}}}.
\label{MA1}
\end{eqnarray}
Here we have denoted:
\begin{eqnarray}
{\cal B}_{1,2}:= b_1 b_2\, {\cal K}^2+\frac{1}{4}\left( b_{1,2}\,c
+b_{2,1}\,d \right)^2,
\label{B12}
\end{eqnarray}
\begin{eqnarray}
{\cal Q}:=&\left(\sqrt{b_1 b_2-c^2}+\sqrt{b_1 b_2-d^2}\right)^2
\left[b_1 b_2 \left(\sqrt{{\cal M}_1}+\sqrt{{\cal M}_2}\right)^2\right.\nonumber\\
&-\left.\frac{1}{4}(b_1-b_2)^2\right]-\left(\sqrt{{\cal B}_1}
-\sqrt{{\cal B}_2}\right)^2.
\label{C}
\end{eqnarray}
The notations ${\cal K}, {\cal M}_1, {\cal M}_2$ are introduced in Eqs.\ (\ref{tildeb1}) and\ (\ref{M}).

Before proceeding to apply the formula\ (\ref{MA1}) to several notorious GSs, we recall the Peres-Simon theorem \cite{Si}: Preservation of non-negativity of the density matrix under 
partial transposition (PT) is a necessary and sufficient 
condition for the separability of any two-mode GS. 
The standard form\ (\ref{st1})--\ (\ref{st2}) of the CM 
of a separable two-mode GS $\hat \rho$ transforms into that 
of its positive partial transpose ${\hat\rho}^{PT}$ just 
by changing the sign of the parameter $d$: $d\rightarrow -d$.  
In view of a lemma proven by Simon \cite{Si}, the two-mode GSs 
with positive $d$ are separable. However, their discord does not vanish \cite{PG,AD} and can be evaluated within our present approach.

\section{Symmetric two-mode Gaussian states}

As a first application of the formula\ (\ref{MA1}), we now consider  
the geometric discord for the noteworthy class of the symmetric 
two-mode GSs. These are defined as states possessing equal marginal purities, i. e., with $b_1=b_2=:b.$  
Notice that definition\ (\ref{B12}) implies the equality ${\cal B}_{1}={\cal B}_{2}=:{\cal B}$. The explicit symplectic eigenvalues of the CM of a symmetric two-mode GS \cite{PT2008},  
\begin{equation}
\kappa_{1}=\sqrt{(b+c)(b+d)},\quad \kappa_{2}=\sqrt{(b-c)(b-d)}, 
\label{ksym}
\end{equation}
allow us first to write $\kappa_{1}^2-\kappa_{2}^2=2 b(c+d),$ and then 
Eq.\ (\ref{B12}) becomes via Eqs.\ (\ref{tildeb2})--\ (\ref{N}):
\begin{eqnarray}
{\cal B}={\cal K}^2\left[b^2+\frac{1}{4}\left(\sqrt{{\cal M}_1}
-\sqrt{{\cal M}_2}\right)^2\left(\sqrt{{\cal N}_1}
-\sqrt{{\cal N}_2}\right)^2\right].
\label{ks} 
\end{eqnarray}
After a routine algebra, the Gaussian Hellinger discord, 
Eq.\ (\ref{gdh}), simplifies to:
\begin{eqnarray}
& D_H({\hat \rho}_{sym})=1-\mathcal A (\hat \rho_{sym}) 
\nonumber \\
& =1-\frac{4(\det {\cal V})^{\frac{1}{4}}}{ \kappa_{1}^{PT}
+\kappa_{2}^{PT}+2(\det {\cal V})^{\frac{1}{4}}
\left( \sqrt{{\cal N}_1}-\sqrt{{\cal N}_2}\right)}.
\label{s1}
\end{eqnarray}
Here we have employed the symplectic eigenvalues of the CM 
of the PT symmetric two-mode GS:
\begin{equation}
\kappa_{1}^{PT}=\sqrt{(b+c)(b-d)}, \quad \kappa_{2}^{PT}
=\sqrt{(b-c)(b+d)}. 
\label{ksympt}
\end{equation}
The quantities  ${\cal N}_1,\; {\cal N}_2$ 
are defined in Eq.\ (\ref{N}). Equation\ (\ref{s1}) 
encompasses both cases $d=\pm|d|$. 

It is remarkable that the obtained maximal affinity
of the symmetric GSs depends only on the symplectic eigenvalues of the state and of its partial transpose (PT). 
 Moreover, for the symmetric GSs with $d=-|d|$ we can compare this dependence with some results for the degree of entanglement. 
 Recall that the symmetric GSs are entangled when the condition  
$\kappa_{2}^{PT}=\sqrt{(b-c)(b-|d|)}<\frac{1}{2}$
is met. Their entanglement of formation could be exactly evaluated 
\cite{G1} and has been found to depend only of the smallest symplectic eigenvalue $\kappa_{2}^{PT}$ of the CM 
of the PT symmetric GS. A similar $\kappa_{2}^{PT}$-dependence 
has been found by the present authors for a Gaussian distance-type 
degree of entanglement defined with the Bures metric \cite{PT2008}.

\section{States with $|d|=c$}

Owing to their experimental applications, the two-mode GSs 
identified by the relation $|d|=c$ are particularly relevant 
when evaluating quantum correlations.  The sign of the parameter 
$d$ splits this case in so far that we discuss two distinct classes 
of states. 

For negative $d=-c$, we deal with the familiar {\em two-mode squeezed 
thermal states} (STSs) \cite{PTH2003}.  According to Simon's separability criterion \cite{Si}, a STS is separable when it meets the condition 
\begin{equation}
\left( b_1-\frac{1}{2}\right) \left( b_2-\frac{1}{2}\right)
-c^2\geqq 0,
\label{sep}
\end{equation}
 and entangled in the opposite case \cite{PTH2001,PTH2003}. 

On the contrary, all the GSs with positive $d=c$ are separable. Specifically, such a state is the partial transpose of a separable STS. Since states with $d=c>0$ can be obtained by mixing two input modes in distinct TSs in a lossless beam splitter, we call them 
{\em mode-mixed thermal states} (MTSs).

According to Eqs.\ (\ref{gdh}) and\ (\ref{MA}), the Gaussian geometric discord simplifies to
\begin{eqnarray}
 D_H ({\hat \rho}_{|d|=c})=\frac{\sqrt{\tilde b_1\tilde b_2}
-\sqrt{\tilde b_1\tilde b_2
-{\tilde c}^2}}{\sqrt{\tilde b_1\tilde b_2}
+\sqrt{\tilde b_1\tilde b_2-{\tilde c}^2}}.
\label{D-sts}
\end{eqnarray}
Our aim is to express the Gaussian discord  
$D_H ({\hat \rho}_{|d|=c})$ in terms of the entries of the CM 
$\cal V$ of the state $\hat \rho$ for both classes of states presented above. We can do this by using our general 
equations\ (\ref{tildeb})--\ (\ref{tildeb4}) and specializing 
the maximal affinity\ (\ref{MA}). The current parametrizations of these states enable us to obtain more insightful formulae.

\subsection{Case $d=-c<0:$ Two-mode squeezed thermal states} 

A STS $\hat \rho_{ST}$ is the result of the action of a two-mode 
squeeze operator  \cite{SC1985}, 
\begin{equation}
\hat S_{12}(r,\phi):=\exp{\left[ \,r \left( {\rm e}^{i\phi} 
\hat a^{\dag}_1 \hat a^{\dag}_2-{\rm e}^{-i\phi} \hat a_1 
\hat a_2 \right) \right] },\nonumber\\
 \left( r>0,\;\; 
\phi\in (-\pi,\pi] \right),
\label{S_12}
\end{equation} 
on a product-thermal state \cite{PTH2003}:
\begin{equation}
 \hat \rho_{ST}=\hat S_{12}(r,\phi)\hat \rho_T(\kappa_1,\kappa_2) 
 \hat S^{\dag}_{12}(r,\phi),
\label{STS}
\end{equation} 
The symplectic eigenvalues of its CM are determined 
by the thermal mean occupancies $\bar n_j$ of the two-mode TS 
$\hat \rho_T(\kappa_1,\kappa_2): \, \kappa_j=\bar n_j+\frac{1}{2},\;
(j=1,2)$. We find the  entries of the standard-form CM ${\cal V}$ depending on the squeeze parameter $r$ as follows \cite{PTH2003}: 
\begin{eqnarray}
&b_{1,2}=\kappa_{1,2}[\cosh(r)]^2
+\kappa_{2,1}[\sinh(r)]^2,\nonumber \\
&c=-d=(\kappa_{1}+\kappa_{2})\cosh(r)\sinh(r).
\label{sts1}
\end{eqnarray}
Insertion of Eq.\ (\ref{sts1}) into Eq.\ (\ref{sep}) allows one
to write explicitly the separability threshold $r_s$ of the STSs
\ (\ref{STS}) \cite{PTH2003}:
\begin{equation}
[\sinh(r)]^2 \leqq [\sinh(r_s)]^2:=\frac{\bar n_1 \bar n_2}
{\bar n_1+\bar n_2+1}.
\label{r_s}
\end{equation}
Any separable two-mode STS fulfils the condition $r \leqq r_s,$
Eq.\ (\ref{r_s}), and, moreover, is  classical, i. e., it has 
a well-behaved Glauber-Sudarshan $P$ representation.  

From the general equations\ (\ref{tildeb})--\ (\ref{tildeb4})  
we get the maximal affinity
\begin{eqnarray}
\mathcal A (\hat \rho_{ST})=\frac{2\left( \sqrt{{\cal N}_1}
+\sqrt{{\cal N}_2}\right)}{\sqrt{{\cal N}_1}+\sqrt{{\cal N}_2}
+\left [\left( \sqrt{{\cal N}_1}+\sqrt{{\cal N}_2}\right)^2
+4c^2\right]^{\frac{1}{2}}}.
\label{maxaff2}
\end{eqnarray}
Then, taking into account the definitions\ (\ref{N}),  
Eq.\ (\ref{maxaff2}) simplifies to
\begin{eqnarray}
\mathcal A (\hat \rho_{ST})=\frac{2 (\kappa_1+\kappa_2)}{\kappa_1
+\kappa_2+[(\kappa_1+\kappa_2)^2
+8c^2(\kappa_1\kappa_2+1/4-\sqrt{\cal D})]^{\frac{1}{2}}},
\label{maxaff2-}
\end{eqnarray}
where the symplectic eigenvalues $\kappa_{1,2}$ of a STS are expressed 
in terms of the standard-form parameters as it follows 
from Eq.\ (\ref{sts1}):
\begin{equation}
\kappa_{1,2}=\frac{1}{2}\left[ \sqrt{(b_1+b_2)^2-4 c^2}
\pm (b_1-b_2)\right].
\label{kappaSTS}
\end{equation}
The maximal affinity\ (\ref{maxaff2-}) becomes 
\begin{equation}
\mathcal A (\hat \rho_{ST})=\frac{2}{\sqrt{\cal X}+1}, \quad
{\cal X}:=1+2\left(\sqrt{\det{\cal V}}+\frac{1}{4}
-\sqrt{{\cal D}}\right)[\sinh(2r)]^2,
\label{X}
\end{equation} 
where the symplectic invariant ${\cal D}$ is written 
in Eq.\ (\ref{tildeb4}). Alternatively, Eq.\ (\ref{X}) can be obtained 
by making direct use of the standard-form CM of the square-root state $(\hat\rho_{ST})_{sr}$, Eq.\ (\ref{sr}), whose entries are: 
\begin{eqnarray}
&\tilde b_{1,2}=\tilde\kappa_{1,2}[\cosh(r)]^2+\tilde\kappa_{2,1}
[\sinh(r)]^2,\nonumber \\
&\tilde c=-\tilde d=(\tilde\kappa_{1}
+\tilde\kappa_{2})\cosh(r) \sinh(r),
\label{sts2}
\end{eqnarray}
with $\tilde\kappa_{1,2}$ given by Eq.\ (\ref{srsymp}).
After some algebra we recover the result\ (\ref{X}) 
that we now specialize to a couple of interesting cases.

Let us first write the geometric discord for {\em symmetric} STSs $(b_1=b_2=:b)$. Equation\ (\ref{X}) leads to an expression similar to Eq.\ (\ref{D-sts}):
\begin{equation}
 D_H({\hat \rho}_{ST})=\frac{b-\sqrt{b^2-{c}^2}}{b+\sqrt{b^2
 -{c}^2}},\qquad (b_1=b_2=:b).
\label{D-sts2}
\end{equation} 
In addition, by applying Eq.\ (\ref{sts1}), we find that 
the geometric discord quantified by Hellinger distance for all 
the symmetric STSs is independent of the degree of mixing and 
has the simple expression
\begin{eqnarray}
 D_H({\hat \rho}_{ST})=[\tanh(r)]^2.
\label{sym}
\end{eqnarray} 
Note that this set of states includes the two-mode squeezed vacuum ones \cite{SC1985}:
\begin{equation}
 \hat \rho_{SV}=\hat S_{12}(r,\phi)|0,0\rangle \langle 0,0|
 \hat S^{\dag}_{12}(r,\phi).
\label{SVS}
\end{equation} 
In view of Eqs.\ (\ref{kappaSTS}) and\ (\ref{sts1}), respectively,
the characterization\ (\ref{sfpSVS}) of these pure states gives
the minimal symplectic eigenvalues $\kappa_1=\kappa_2=\frac{1}{2}$
and then the standard-form parameters $b=\frac{1}{2}\cosh(2r),\, 
c=\frac{1}{2}\sinh(2r).$

Another simplification arises for any STS whose CM has a single minimal symplectic eigenvalue: $\kappa_2=\frac{1}{2}$. Equation\ (\ref{X}) 
then gives the Hellinger discord
\begin{equation}
D_H({\hat \rho}_{ST})=\frac{\{(\bar n_1+1)[\sinh(2r)]^2+1\}^{1/2}-1}
{\{(\bar n_1+1)[\sinh(2r)]^2 +1\}^{1/2}+1},
\quad \left( \kappa_2=\frac{1}{2}\right).
\end{equation}

As noticed earlier, the original geometric discord defined 
with the Hilbert-Schmidt metric proved to display 
some inconveniences \cite{P} related to its property 
of non-contractivity.  Quite recently, a modification 
of the Hilbert-Schmidt norm for continuous-variable systems was addressed in Ref. \cite{adesso2} in order to prevent 
its dependence on the global purity of the states involved. The result reported in Eq. (B.6) of Ref. \cite{adesso2} for two-mode STSs can now 
be compared to our Eq.\ (\ref{D-sts}). The two measures share the same functional form but with the significant difference of having the parameters $\tilde {b}_1, \tilde{b}_2,  \tilde{c}
= |\tilde{d}|$  replaced in Ref. \cite{adesso2} by the corresponding entries of the CM ${\cal V}$. Only for symmetric STSs 
our Eq.\ (\ref{X}) and Eq. (B.6) of Ref.\cite{adesso2} give 
the same result, namely, Eq.\ (\ref{sym}). For asymmetrical STSs, 
the difference between the two expressions of the geometric discord 
are determined by the asymmetry of the modes and, in general, 
we found them very close. This can be seen in Fig. 1, 
where the monotonic behaviour of two geometric measures of discord with the degree of squeezing is displayed by a non-symmetric example. 

\begin{figure}[t,h]
\center
\includegraphics[width=6cm]{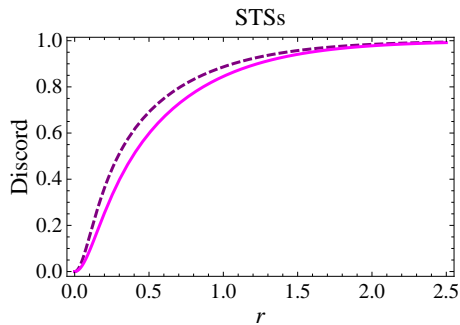}
\caption{(Color online) Hellinger discord (purple dotted plot), 
Eq.\ (\ref{D-sts}), and the rescaled Hilbert-Schmidt one (magenta line plot), Eq.(B.6) of Ref.\cite{adesso2}, versus the squeeze parameter $r$ 
for a non-symmetric STS  with the symplectic eigenvalues 
$\kappa_1=0.5$ and $\kappa_2=20.5$.}
\end{figure}

\subsection{Case $d=c>0:$ Mode-mixed thermal states}

It is well known that a beam splitter mixes two incident modes 
to produce two outgoing ones. Recall that the optical interference 
of two modes in a reversible, lossless beam splitter is described 
by a mode-mixing operator \cite{Bonny}:
\begin{equation}
\hat M_{12}(\theta,\phi):=\exp{\left[ \frac{\theta}{2}
\left( {\rm e}^{i\phi} \hat a_1  \hat a^{\dag}_2-{\rm e}^{-i\phi} 
\hat a^{\dag}_1  \hat a_2 \right) \right] }.
\label{M_12}
\end{equation} 
As a matter of fact, this is a $SU(2)$ displacement operator
\cite{Radcliffe, Arecchi} written employing the Jordan-Schwinger
two-mode bosonic realization of angular momentum \cite{BS, Ulf}. 
Its parameters are the spherical polar angles $\theta$ and 
$\phi: \theta\in [0,\pi],\; \phi\in (-\pi,\pi]$.  The co-latitude 
$\theta$ determines the intensity transmission and reflection coefficients of the device, which are  
$T=\left[ \cos \left( \frac{\theta}{2}\right) \right]^2$
and, respectively,
$R=\left[ \sin \left( \frac{\theta}{2}\right) \right]^2.$ 

When choosing an asymmetrical two-mode TS as input to a beam splitter, 
then we get an emerging MTS as output:
\begin{equation}
\hat \rho_{MT}=\hat M_{12}(\theta,\phi)\hat \rho_T(\kappa_1,\kappa_2)
\hat M^{\dag}_{12}(\theta,\phi), \quad (\kappa_1 >\kappa_2).
\label{MTS}
\end{equation} 
The standard-form entries of the output CM are found to be:
\begin{eqnarray}
&b_{1,2}=\kappa_{1,2}\left[ \cos \left( \frac{\theta}{2}\right)
\right]^2+\kappa_{2,1}\left[ \sin \left( \frac{\theta}{2}\right) \right]^2,\nonumber \\
&c=d=(\kappa_{1}-\kappa_{2})\cos \left( \frac{\theta}{2}\right)
\sin \left(\frac{\theta}{2}\right).
\label{m1}
\end{eqnarray}
As already mentioned, any MTS $\hat \rho_{MT}$  is a separable two-mode GSs with $d=c>0$ and, consequently, it is related to a separable STS 
by partial transposition \cite{Si}. Furthermore, any MTS 
$\hat \rho_{MT}$ is a classical state, i. e., it possesses a well-behaved 
Glauber-Sudarshan $P$ representation. 
The general equations\ (\ref{tildeb})--\ (\ref{tildeb4}) yield 
the maximal affinity
\begin{eqnarray}
\mathcal A (\hat \rho_{MT})=\frac{2\left( \sqrt{{\cal M}_1}
+\sqrt{{\cal M}_2}\right)}{\sqrt{{\cal M}_1}+\sqrt{{\cal M}_2}
+\left[ \left(\sqrt{{\cal M}_1}+\sqrt{{\cal M}_2}\right)^2
+4c^2\right]^{\frac{1}{2}}}.
\label{maxaff1}
\end{eqnarray}
By use of Eq.\ (\ref{M}), Eq.\ (\ref{maxaff1}) simplifies to
\begin{eqnarray} 
\mathcal A (\hat \rho_{MT}) =\frac{2(\kappa_1-\kappa_2)}
{\kappa_1-\kappa_2+\left[ (\kappa_1-\kappa_2)^2
+8c^2\left(\kappa_1\kappa_2-\frac{1}{4}-\sqrt{\cal D}\right)
\right]^{\frac{1}{2}}}.
\label{maxaff1+}
\end{eqnarray}
From Eq.\ (\ref{m1}) we get the symplectic eigenvalues $\kappa_{1,2}$ 
of the CM of a MTS expressed in terms of its standard-form entries: 
\begin{equation}
\kappa_{1,2}=\frac{1}{2}\left[ b_1+b_2 \pm  \sqrt{(b_1-b_2)^2
+4 c^2}\right].
\label{kappaMTS}
\end{equation}
The maximal affinity\ (\ref{maxaff1+}) becomes 
\begin{equation}
\mathcal A (\hat \rho_{MT})=\frac{2}{\sqrt{\cal Y}+1}, \quad
{\cal Y}:=1+2\left(\sqrt{\det{\cal V}}-\frac{1}{4}
-\sqrt{{\cal D}}\right)[\sin(\theta)]^2.
\label{Y}
\end{equation}

\section{Discussion and conclusions}

In view of the analysis in Section 4, an examination of the consistency between the Gaussian discord\ (\ref{gdh}) evaluated via Eq.\ (\ref{MA})  and the original exact one \cite{PG,AD} becomes a necessity. 
To make our presentation as simple as possible, we choose to deal with symmetric two-mode GSs ($b_1=b_2=:b$) having $|d|=c$.  
In this particular case, the standard-form parameter $b$ is related 
to the total mean photon number $\langle \hat N \rangle$ of the state:
\begin{equation}
b=\frac{1}{2}\left(\langle \hat N \rangle +1 \right). 
\label{b}
\end{equation}
 By specializing the Gaussian discord 
of Refs.\cite{PG,AD}, we get the formula: 
\begin{equation}
{\mathcal D}(\hat \rho_{|d|=c})=h(b)-h(\kappa_1)-h(\kappa_2)+h(y).
\label{PA}
\end{equation}
Here $h(x)$ is the entropic function
\begin{equation}
h(x):=\left( x+\frac{1}{2}\right)
\ln\left (x+\frac{1}{2}\right)
-\left( x-\frac{1}{2}\right)
\ln\left (x-\frac{1}{2}\right), \quad \left( x\geqq \frac{1}{2}\right),
\label{h}
\end{equation}
$\kappa_{1,2}$ are the symplectic eigenvalues of the CM,  
and  $\;y:=b-\frac{c^2}{b+\frac{1}{2}}$. 
The states being symmetric,  the quantum mutual information\ (\ref{MI}) specializes to
\begin{equation}
{\mathcal I}(\hat \rho_{|d|=c})=2 h(b)-h(\kappa_1)-h(\kappa_2).
\label{MI1}
\end{equation}
As seen from Eq.\ (\ref{PA}),  states whose CMs differ only by the sign of the parameter $d$ have different amounts of quantum correlations because they have different symplectic eigenvalues. However, they have the same amount of classical correlations: 
\begin{equation}
{\mathcal C}(\hat \rho_{|d|=c}):={\mathcal I}(\hat \rho_{|d|=c})-{\mathcal D}(\hat \rho_{|d|=c})=
h(b)-h(y).
\label{CC}\end{equation}
According to Eq.\ (\ref{sep}), for $b-c\geqq\frac{1}{2},$ 
all the symmetric states with $d=\pm c$ are separable.  
In the absence of any entanglement, their geometric Gaussian discord measures all the other Gaussian correlations (classical and quantum). Correlations of such states are described in Fig. 2 for a fixed value 
of their common purity,
$${\rm Tr} \left[(\hat \rho)^2 \right]
=\left[ \det(2{\cal V})\right]^{-\frac{1}{2}}
=\frac{1}{4\left( b^2-c^2 \right)}.$$ 

\begin{figure*}[h]
\center
Separable states with $|d|=c$

\includegraphics[width=7cm]{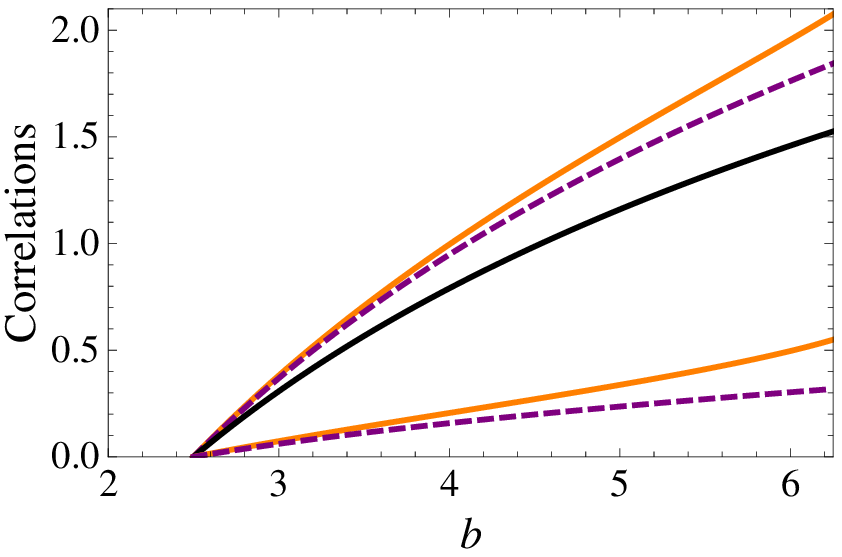}
\includegraphics[width=7cm]{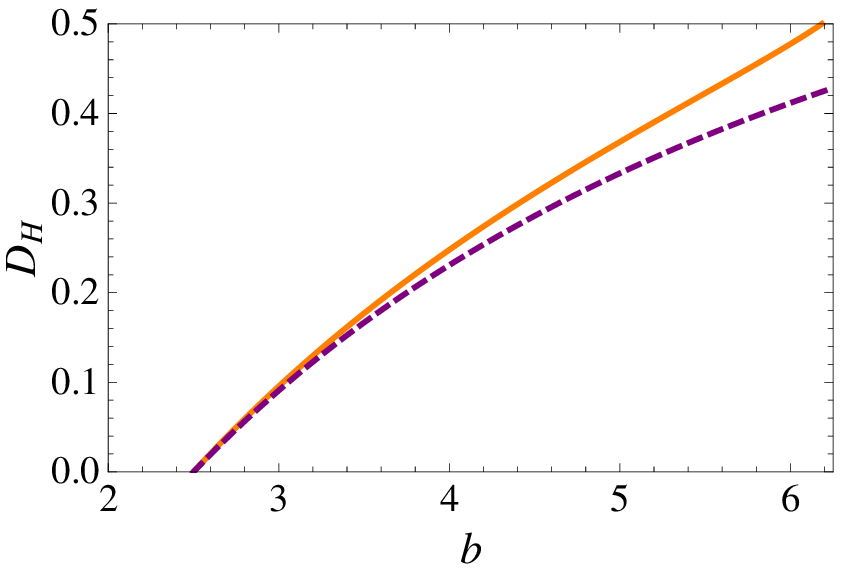}
\caption{ (Color online) We compare the aspect of the geometric Hellinger discord, Eq.\ (\ref{s1}), (right panel), and two measures of Gaussian correlations (left panel), when describing the same sets of separable states. The full line curves represent MTSs, 
$(d=c>0)$, corresponding to  $b^2-c^2=6.25$ (fixed purity). Separable STSs,  $(d=-c<0),$ are shown by purple 
dashed curves at the same value of the purity. The lower curves in the left panel represent the original Gaussian discord, Eq.\ (\ref{PA}), the upper ones are the mutual information, Eq.\ (\ref{MI1}), while the full black plot represents the classical correlations, Eq.\ (\ref{CC}). Note that symmetric STSs and MTSs possess the same amount of classical correlations at the same purity.}
\end{figure*}

It is rather intriguing that the Gaussian discord of the states with $d=c>0$, i. e., MTSs, turns out to be larger than that of the corresponding ones with $d=-c<0$, which are separable STSs. This feature is common to both Gaussian measures we have used: the geometric 
Hellinger discord, Eq.\ (\ref{s1}), (right panel) and the original one (lower curves on the left panel). Remark that in this case the graphs 
of the quantum mutual information (upper curves on the left panel) are just shifted with respect to the discord plots by the same amount 
of classical correlations, Eq.\ (\ref{CC}).
Figure 2 exhibits another important fact too: the two Gaussian
discords\ (\ref{s1}) and\ (\ref{PA}) have a similar behaviour 
with respect to the parameter $b$.  This can be interpreted as an expression of their consistency.

\begin{figure}[h]
\center
\includegraphics[width=7cm]{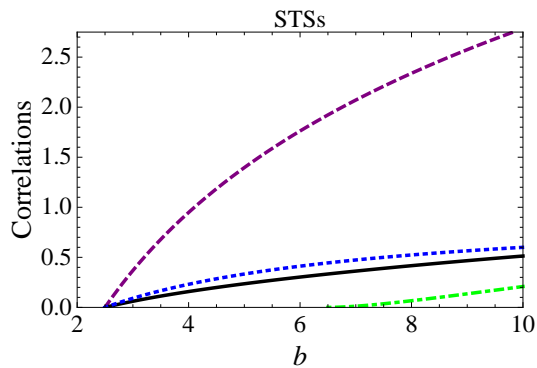}
\caption{(Color online) Measures of Gaussian correlations for a symmetric STS with $\sqrt{\det {\cal V}}=b^2-c^2=6.25$ versus the parameter $b$. The dashed purple curve represents the quantum mutual information, Eq.\ (\ref{MI1}) with $\kappa_1=\kappa_2
=\sqrt{b^2-c^2}$. The black full line plot represents the original Gaussian discord, Eq.\ (\ref{PA}), the blue dotted line is the Hellinger discord, Eq.\ (\ref{D-sts2}), and 
the dashed-dotted green curve is the entanglement of formation, Eq.\ (\ref{EF}).}
\end{figure}

We now examine the influence of entanglement on both Gaussian
discords. Let us thus consider a symmetric STS. It is separable 
for $\kappa^{PT}=b-c\geqq\frac{1}{2}$ and entangled
for $\kappa^{PT}<\frac{1}{2}$. 
In Fig. 3 we plot both discords\ (\ref{s1}) and\ (\ref{PA}) 
for a STS with $b^2-c^2=6.25$ as depending of the parameter $b$. 
The state is entangled for $b>6.5$ and its exact entanglement 
of formation \cite{G1},
\begin{equation}
E_f(\hat \rho_{ST})=h(z),\qquad z:=\frac{(b-c)^2
+\frac{1}{4}}{2(b-c)},
\label{EF}
\end{equation}
is also shown in Fig. 3. We first notice the smooth monotonic increasing of the two measures of Gaussian discord, 
Eqs.\ (\ref{PA}) and\ (\ref{D-sts2}). Both plots are consistent
and seem to be totally insensitive to the absence or presence 
of entanglement.  

To sum up, in this paper we have considered the Hellinger 
metric as a measure of quantum discord for two-mode Gaussian 
states. As any other distance used to define a geometric Gaussian discord, the Hellinger one is a measure of the total amount of correlations, providing just an upper bound of the exact geometric discord. We have reviewed some useful properties of the affinity, 
which is a close relative of the Uhlmann fidelity. It is interesting 
to note that almost simultaneously with launching fidelity 
as a new tool in quantum information processing 
in Ref. \cite{Jozsa}, affinity was analyzed as a possible quantifier 
of the information flow in and out of a black hole \cite{Al}. It was 
its clear meaning  expressed in our Eq.\ (\ref{fi1}) that gave prominence to fidelity  as a measure of closeness between quantum states. However, we here have shown that the recently reconsidered affinity \cite{Luo1,Luo3} and the more popular fidelity share 
some  useful properties for defining distance-type measures 
of quantum properties such as correlations. 
In this paper we have defined and exactly evaluated a geometric discord based on the maximal affinity between a given two-mode GS 
and the whole set of two-mode product GSs. The general analytic formula we have derived has easily been specialized to symmetric 
GSs. A detailed analysis is devoted to a pair of classes 
of two-mode GSs that have insightful parametrizations and at the same time can readily be prepared: the squeezed thermal states 
and the mode-mixed thermal ones. The latter are separable states 
and have been compared with the appropriate separable STSs, 
in order to evaluate their total amounts of correlations in the absence of entanglement by means of the Hellinger discord. We have chosen symmetric states 
from both classes and examined the variation of their Gaussian discords with the mean total number of photons at a fixed global purity. 
Comparison of the Hellinger discord\ (\ref{s1}) with the originally defined one\ (\ref{PA}) indicates consistency by inducing the same ordering of all the Gaussian correlations.

\ack{This work was supported  by the Romanian National Authority for Scientific Research (CNCS – UEFISCDI) through Grant No.~PN-II-ID-PCE-2011-3-1012 and by the University of Bucharest through Grant No. UB 898/2013.}

\appendix  
\section{Relations between the covariance matrices $\cal V$ 
and $\tilde {\cal V}$ for some types of two-mode Gaussian states}

In the body of the paper use is made of the explicit relations between the entries of the CMs $\cal V$ and $\tilde {\cal V}$.
A first example is the product of two single-mode GSs: 
$\hat \rho=\hat \rho_1\otimes\hat \rho_2$. The associate 
square-root state\ (\ref{sr}) is the product GS
$(\hat \rho)_{sr}=(\hat \rho_1)_{sr}\otimes(\hat \rho_2)_{sr}$. 
Both product states have block-diagonal CMs: 
\begin{eqnarray}
{\mathcal V}_p=\left(
\begin{array}  {cc}
{\mathcal V}_{p1}      & 0\\ 
0 & {\mathcal V}_{p2} 
\end{array}
\right),\quad \mathcal V_{pj}=\left(
\begin{array}  {cc}
\sigma^{(j)}_{11} & \sigma^{(j)}_{12}\\ 
\sigma^{(j)}_{12}& \sigma^{(j)}_{22} 
\end{array}
\right),\quad (j=1,2),
\label{st3}
\end{eqnarray}
\begin{eqnarray}
\tilde{\mathcal V}_p=\left(
\begin{array}  {cc}
\tilde{\mathcal V}_{p1}     & 0\\ 
0 & \tilde{\mathcal V}_{p2}
\end{array}
\right),\quad \tilde{\mathcal V}_{pj}=\left(
\begin{array}  {cc}
\tilde{\sigma}^{(j)}_{11} & \tilde{\sigma}^{(j)}_{12}\\ 
\tilde{\sigma}^{(j)}_{12}& \tilde{\sigma}^{(j)}_{22} 
\end{array}
\right),\quad (j=1,2).
\label{srst3}
\end{eqnarray}
Here the symmetric $2\times 2$ matrices $\mathcal V_{pj}$ and 
$\tilde{\mathcal V}_{pj}$ are CMs of one-mode displaced squeezed thermal states in an appropriate parametrization \cite{P92,PT93a,PTH2001}. Accordingly,
\begin{eqnarray}
&\sigma^{(j)}_{11}=\eta_j [\cosh(2 r_j)+\cos (\varphi_j)\sinh(2 r_j)],\nonumber\\
&\sigma^{(j)}_{22}=\eta_j [\cosh(2 r_j)-\cos (\varphi_j)\sinh(2 r_j)],\nonumber\\
&\sigma^{(j)}_{12}=\eta_j\sin(\varphi_j)\sinh(2 r_j),
\label{prod}
\end{eqnarray}
where $\eta_j$ are the symplectic eigenvalues of the CM\ (\ref{st3}), while $r_j$ denote the one-mode positive squeeze parameters and 
$\varphi_j$ the corresponding squeeze angles.  
Taking account of Eqs.\ (\ref{u}),\ (\ref{srsymp}),\ (\ref{usr}), 
and\ (\ref{sr}), we get the CM $\tilde{\cal V}_p$, 
Eq.\ (\ref{srst3}). Its entries are found to have the same expressions and parameters as in Eqs.\ (\ref{prod}), except for the substitution 
$\eta_j\rightarrow \tilde{\eta}_j:=\eta_j+\sqrt{\eta^2_j
-\frac{1}{4}}$. Hence,
\begin{equation}
\tilde {\mathcal V}_{pj}
={\mathcal V}_{pj} (\tilde{\eta}_j,r_j,\varphi_j),\qquad (j=1,2).
\end{equation}

The second case we deal with is the CM $\tilde{\mathcal V}$ corresponding to a scaled standard-form CM ${\mathcal V}$. 
Using Eq.\ (\ref{tildeV}) we find after some algebra that the CM  
$\tilde{\mathcal V}$ has the structure displayed 
by Eqs.\ (\ref{st1}) and\ (\ref{st2}) with the following entries:
\begin{eqnarray}
\tilde b_{1,2}\tilde s_{1,2}=\left[b_{1,2} {\cal L}-b_{2,1}(b_1b_2-c^2)\right]\frac{1}{4\kappa_1\kappa_2 {\cal K}} s_{1,2}, \nonumber\\
\tilde b_{1,2} \frac{1}{\tilde s_{1,2}}=\left[b_{1,2} {\cal L}-b_{2,1}(b_1b_2-d^2)\right]\frac{1}{4\kappa_1\kappa_2 {\cal K}} \; 
\frac{1}{s_{1,2}}, \nonumber\\
\tilde c\sqrt{\tilde s_1\tilde s_2} =\left[c \,{\cal L}+d \,(b_1b_2-c^2)\right]\frac{1}{4\kappa_1\kappa_2 {\cal K}}\;\sqrt{s_1s_2}, \nonumber\\
\tilde d\frac{1}{\sqrt{\tilde s_1\tilde s_2}} =\left[d\, {\cal L}+c \,(b_1b_2-d^2)\right]\frac{1}{4\kappa_1\kappa_2 {\cal K}}\;\frac{1}{\sqrt{s_1s_2}}.
\label{tildeb}
\end{eqnarray}
In Eq.\ (\ref{tildeb}) we have introduced two symplectic invariants: 
\begin{eqnarray}
&{\cal K}:=\kappa_1\sqrt{\kappa_2^2-\frac{1}{4}}
+\kappa_2\sqrt{\kappa_1^2-\frac{1}{4}},
\nonumber\\
&{\cal L}:=4\sqrt{\det \mathcal V \det \tilde{\mathcal V}}
=4 \kappa_1\kappa_2\tilde \kappa_1\tilde \kappa_2.
\label{tildeb1}
\end{eqnarray}
It is useful to note a factorization of the first one:
\begin{eqnarray}
{\cal K}= \frac{1}{2}\left(\sqrt{{\cal M}_1}+\sqrt{{\cal M}_2}\right)
\left(\sqrt{{\cal N}_1}+\sqrt{{\cal N}_2}\right),
\label{tildeb2}
\end{eqnarray}
where we have employed the non-negative products
\begin{equation}
{\cal M}_1:=\left(\kappa_1-\frac{1}{2}\right)\left(\kappa_2+\frac{1}{2}\right),\quad {\cal M}_2:=\left(\kappa_1+\frac{1}{2}\right)\left(\kappa_2
-\frac{1}{2}\right),
\label{M} 
\end{equation}
\begin{equation}
{\cal N}_1:=\left(\kappa_1+\frac{1}{2}\right)\left(\kappa_2
+\frac{1}{2}\right),\quad {\cal N}_2:
=\left(\kappa_1-\frac{1}{2}\right)\left(\kappa_2-\frac{1}{2}\right),
\label{N}
\end{equation}
satisfying the obvious identity ${\cal M}_1 {\cal M}_2
={\cal N}_1 {\cal N}_2={\cal D}$. Here ${\cal D}$ denotes 
the determinant
\begin{eqnarray}{\cal D}:=\det \left({\cal V}+\frac{i}{2} J\right)
=\det{\cal V}-\frac{1}{4}(b_1^2+b_2^2+2 c\, d)+\frac{1}{16},
\label{tildeb4}
\end{eqnarray} 
which is a basic symplectic invariant of the state $\hat \rho$.

From Eqs.\ (\ref{tildeb}) we learn that the CM $\tilde{\mathcal V}$ corresponding to a scaled standard-form CM 
${\mathcal V}$ is in a scaled standard form as well.  
More interesting, we find that for a state without local squeezings, namely, $s_1=s_2=1$ in Eq.\ (\ref{st2}), the CM $\tilde{\mathcal V}$ 
of the square-root state is in general scaled: 
$\tilde s_1\neq 1,\; \tilde s_2\neq 1$. The only exceptions are 
the two families of GSs with $|d|=c$ discussed in Sec. 7: STSs and MTSs. 
Equations\ (\ref{tildeb}) give then 
$s_1=s_2=1\Longrightarrow \tilde s_1=\tilde s_2=1$.


\section*{References}
\begin{thebibliography}{50}
\bibitem{OZ} Ollivier H and Zurek W H 2001
{\em Phys. Rev. Lett.} {\bf 88} 017901
\bibitem{HV} Henderson L and Vedral V 2001
{\em J. Phys. A} {\bf 34} 6899 
\bibitem{Bennett} Bennett C H, DiVincenzo D P, Smolin J A, and  Wootters W K 1996 {\em Phys. Rev. A}  {\bf 54} 3824 
\bibitem{Huang1} Huang Y 2014 {\em New J. Phys.} {\bf 16}  033027
\bibitem{alber} Ali M,  Rau A R P, and Alber G 2010 
{\em Phys. Rev. A } {\bf 81} 042105 
\bibitem{Huang2} Huang Y 2013 {\em Phys. Rev. A } {\bf 88} 014302
\bibitem{PG} Giorda P and Paris M G A 2010 {\em Phys. Rev. Lett.} {\bf 105} 020503 
\bibitem{AD} Adesso G and Datta A 2010 Phys. Rev. Lett. {\bf 105} 030501 
\bibitem{PSBCL} Pirandola S, Spedalieri G, Braunstein S L, Cerf N J, and Lloyd S 2014 {\em Phys. Rev. Lett.} {\bf 113} 140405
\bibitem{modi1} Modi K, Brodutch A , Cable H,  Paterek T, 
and  Vedral V 2012 {\em Rev. Mod. Phys.} {\bf 84} 1655 
\bibitem{modi2} Modi K,  Paterek T, Son W, Vedral V, and Williamson M 2010 {\em Phys. Rev. Lett.} {\bf 104} 080501 
\bibitem{DVB} Daki\'c B, Vedral V, and  Brukner C 2010 
{\em Phys. Rev. Lett. }{\bf 105} 190502 
\bibitem{Luo2} Luo S and  Fu S 2010 {\em Phys. Rev. A} {\bf 82} 034302 
\bibitem{Hil} Hillery M 1987 {\em Phys. Rev. A} {\bf 35} 725; 1989 
{\em ibid.} {\bf 39} 2994 
\bibitem{PVK} Vedral V, Plenio M B, Rippin M A, 
and Knight P L 1997 {\em Phys. Rev. Lett.} {\bf 78} 2275 
\bibitem{Dod} Dodonov V V, Man'ko O V, Man'ko V I, 
and  W\"{u}nsche A 2000 {\em  J. Mod. Opt.}  {\bf 47} 633 
\bibitem{PTH02} Marian P, Marian T A, and Scutaru H 2002
{\em  Phys. Rev. Lett.}  {\bf 88} 153601 
\bibitem{Luo1} Luo S and Zhang Q 2004 {\em Phys. Rev. A} {\bf 69} 032106 
\bibitem{PT2013} Marian P and Marian T A 2013
{\em Phys. Rev. A }{\bf 88} 012322 
\bibitem{BGPT} Boca M, Ghiu I, Marian P, and Marian T A 2009 
{\em Phys. Rev. A } {\bf 79} 014302 
\bibitem{GBPT} Ghiu I,  Bj\"{o}rk G, Marian P, and Marian T A 2010 {\em  Phys. Rev. A } {\bf 82} 023803 
\bibitem{PHH} Piani M,  Horodecki P, and Horodecki R 2008 
{\em Phys. Rev. Lett.} {\bf 100} 090502 
\bibitem{P} Piani M 2012 {\em  Phys. Rev. A } {\bf 86} 034101 
\bibitem{Luo3}  Chang L and  Luo S 2013 {\em  Phys. Rev. A} 
{\bf 87} 062303 
\bibitem{adesso3} Girolami D, Tufarelli T, Adesso G 2013 {\em Phys. Rev. Lett.} {\bf 110} 240402
\bibitem{SO} Spehner D and  Orszag M 2013 {\em New J. Phys.} 
{\bf 15} 103001 
\bibitem{adesso1} Adesso G and Girolami D 2011  
{\em Int. J. Quant. Inf.} {\bf 9} 1773 
\bibitem{adesso2} Tufarelli T,  MacLean T, Girolami D, Vasile R, 
and Adesso G 2013 {\em  J. Phys. A:  Math. Theor.} {\bf 46} 275308 
\bibitem{PT2012} Marian P and Marian T A 2012 {\em Phys. Rev. A} 
{\bf 86} 022340
\bibitem{B} Ballentine L E 1986 {\em Am. J. Phys.} {\bf 54} 883 
\bibitem{Fuchs} Fuchs C A 1995 {\em  PhD thesis, University 
of New Mexico)} (quant-ph/9601020/1996)
\bibitem{FC} Fuchs C A and Caves C M 1995 {\em Open Syst. Inform.
Dynamics} {\bf 3} 345
\bibitem{barn1} Barnum H, Caves C M, Fuchs C A,  Jozsa R, and  Schumacher B 1996 {\em Phys. Rev. Lett.} {\bf 76} 2818 
\bibitem{barn2} Barnum H, Fuchs C A,  Jozsa R, and  Schumacher B 
1996 {\em Phys. Rev. A} {\bf 54} 4707 
\bibitem{Uhl} Uhlmann A 1976 {\em Rep. Math. Phys.} {\bf 9} 273
\bibitem{Bures} Bures D 1969 {\em Trans. Am. Math. Soc.} 
{\bf 135} 199
\bibitem{BC} Braunstein S L and Caves C M 1994 
{\em Phys. Rev. Lett.} {\bf 72} 3439 
\bibitem{Jozsa} Jozsa R 1994 {\em J. Mod. Opt.} {\bf 41} 2315 
\bibitem{Mend} Mendon\c{c}a P E M F, Napolitano R d J,
Marchiolli M A, Foster C J, and Liang Y - C 2008 {\em Phys. Rev. A} 
{\bf 78} 052330
\bibitem{Hol}Holevo A S 1972 {\em  Theor. Math. Phys.} {\bf 13} 1071 (english version)
 \bibitem{FG} Fuchs C A and van de Graaf J 1999 
{\em IEEE Trans. Inf. Theory} {\bf 45} 1216 
\bibitem{Breuer} Wi$\beta$mann S, Leggio B , and Breuer H - P 2013 {\em Phys. Rev. A} {\bf 88} 022108 
\bibitem{BL} Braunstein S L  and  Van Loock P 2005 {\em Rev. Mod. Phys.} {\bf 77} 513 
\bibitem{WPGCRSL} Weedbrook C, Pirandola S, 
Garc\'{\i}a-Patr\'{o}n R, Cerf N J, Ralph T C, Shapiro J H, 
and Lloyd S 2012 {\em  Rev. Mod. Phys.} {\bf 84} 621 
\bibitem{QC3} Adesso G, Ragy S, Lee A R 2014 {\em Open Syst. Inf. Dyn.}  {\bf 21} 1440001 
\bibitem{W} Williamson J 1936 {\em Amer. J. Math. } {\bf 58} 141 
\bibitem{PTH2003} Marian P, Marian T A, and Scutaru H 2003 
{\em Phys. Rev. A} {\bf 68} 062309 
\bibitem{PT2008} Marian P and Marian T A 2008 {\em Phys. Rev. A} 
{\bf 77} 062319 
\bibitem{Si} Simon R 2000 {\em Phys. Rev. Lett.} {\bf 84} 2726 
\bibitem{G1} Giedke G,  Wolf M M, Kr\"{u}ger O, Werner R F,
and Cirac J I 2003 {\em  Phys. Rev. Lett.} {\bf 91} 107901 
\bibitem{PTH2001} Marian P, Marian T A, and  Scutaru H 2001  
{\em J. Phys. A:  Math. Gen.} {\bf 34} 6969
\bibitem{SC1985} Schumaker B L and Caves C M 1985 {\em  Phys. Rev. A} 
{\bf 31} 3093
\bibitem{Bonny} Schumaker B L 1986 {\em Phys. Reports} {\bf 135} 317
\bibitem{Radcliffe} Radcliffe J M 1971 {\em  J. Phys. A: Gen. Phys.}
{\bf 4} 313 
\bibitem{Arecchi} Arecchi F T, Courtens E, Gilmore R, 
and Thomas H 1972 {\em Phys. Rev. A} {\bf 6} 2211 
\bibitem{BS} Leonhardt U 1993 {\em  Phys. Rev. A} {\bf 48} 3265 
\bibitem{Ulf} Leonhardt U 2010 {\em Essential Quantum Optics. 
From Quantum Measurements to Black Holes}, (Cambridge University Press, Cambridge, UK)
\bibitem{Al} Albrecht A 1994 {\em Phys. Rev. D} {\bf 50} 2744 
\bibitem{P92} Marian P 1992 {\em Phys. Rev. A} {\bf 45} 2044 
\bibitem{PT93a} Marian P and Marian T A 1993 {\em Phys. Rev. A} 
{\bf  47} 4474 







\end{thebibliography}
\end{document}